\documentclass{ws-ijmpe}

\begin{document}

\markboth{R. \'Alvarez-Rodr\'{\i}guez, A.S. Jensen, D.V. Fedorov and
  E. Garrido} {Momentum distributions from three-body decaying $^{9}$Be
  and $^{9}$B resonances}

\catchline{}{}{}{}{}

\title{MOMENTUM DISTRIBUTIONS FROM THREE-BODY DECAYING $^{9}$Be AND
  $^9$B RESONANCES}

\author{\footnotesize R. \'ALVAREZ-RODR\'IGUEZ} 

\address{Grupo de F\'{\i}sica Nuclear, Departamento de F\'{\i}sica
  At\'omica, Molecular y Nuclear\\ Universidad Complutense de Madrid,
E-28040 Madrid, Spain,\\ raquel.alvarez@fis.ucm.es}

\author{\footnotesize A.S. JENSEN, D.V. FEDOROV}

\address{Department of Physics and Astronomy, University of Aarhus,
  DK-8000 Aarhus C, Denmark}

\author{\footnotesize E. GARRIDO}

\address{Instituto de Estructura de la Materia, CSIC, Serrano 123,
  E-28006 Madrid, Spain}

\maketitle

\begin{history}
\received{(received date)}
\revised{(revised date)}
\end{history}

\begin{abstract}
  The complex-rotated hyperspherical adiabatic method is used to study
  the decay of low-lying $^9$Be and $^9$B resonances into $\alpha$,
  $\alpha$ and $n$ or $p$. We consider six low-lying resonances of
  $^9$Be ($1/2^\pm$, $3/2^\pm$ and $5/2^\pm$) and one resonance of
  $^9$B ($5/2^-$) to compare with. The properties of the resonances at
  large distances are decisive for the momentum distributions of the
  three decaying fragments. Systematic detailed energy correlations of
  Dalitz plots are presented.
\end{abstract}

\section{Introduction}

The structure of the $^9$Be nucleus has been considered as a prototype
of cluster structure in nuclei. It has been described theoretically by
many different cluster models and several experiments have been
performed in order to understand its structure and decay
mechanisms. Much effort has been devoted to the $5/2^-$ state due to
its astrophysical importance in the formation of $^9$Be, $^{12}$C and
heavier elements in stellar nucleosynthesis\cite{pap07,alv08,bur10}.
$^9$Be and $^9$B are mirror nuclei that decay into $\alpha\alpha n$
and $\alpha \alpha p$ respectively. Therefore their structures are
expected to be very similar.  We present here the results of
two-dimensional energy correlations after the three-body decay of
several $^9$Be resonances. Moreover the same kind of results for the
$5/2^-$-resonance of $^9$B are shown in order to show the
similarities.

\section{Theoretical Description Of Three-Body Resonances}

We employ here the complex-scaled hyperspherical adiabatic expansion
method to solve the Faddeev equations which describe our three-body
system\cite{nie01}. The angular part of the Hamiltonian is first
solved keeping fixed the value of the hyperradius $\rho$. Its
eigenvalues serve as effective potentials while the eigenfunctions,
$\Phi_{nJM}$ are used as a basis to expand the total wave-function
$\Psi^{JM} = \frac{1}{\rho^{5/2}}\sum_n f_n (\rho) \Phi_{nJM}
(\rho,\Omega)\;.  $ The $\rho$-dependent expansion coefficients, $f_n
(\rho)$, are the hyperradial wave functions obtained from the coupled
set of hyperradial equations\cite{nie01}.

We consider a two-body interaction able to reproduce the low-energy
scattering properties of the two different pairs of particles in our
three-body system. Ali-Bodmer $\alpha-\alpha$ potential\cite{ali66}
and Coulomb potential between $\alpha$-particles are considered. The
$\alpha$-nucleon interaction is taken from Cobis et
al\cite{cob97}. The $^{9}$Be- and $^9$B-resonances are of three-body
character at large-distances, where they decay into two
$\alpha$-particles and one neutron or proton, but this is not
necessarily the case at short-distances.\cite{gar10} We use the
three-body model at all distances because the decay properties only
require the proper description of the emerging three particles. A
three-body short-range potential of the form $V_{3b} =
S\exp(-\rho^2/b^2)$ is included to adjust the corresponding
small-distance part of the effective potential. The correct resonance
energies, which are all-decisive for decay details as evident in the
probability for tunneling through a barrier, are then correctly
reproduced.

The eigenvalues of the angular Hamiltonian for fixed hyperradius,
serve as adiabatic potentials\cite{alv10}. Each of them correspond to
a specific combination of quantum numbers, i.e., partial-wave momenta
between two particles in each Jacobi system. Usually few angular
eigenvalues are needed for achieving convergence. At small distances
the potentials have wells that support the bound states and
resonances.

\section{Energy Distributions}

The energy distribution is the probability for finding a given
particle at a given energy. It can be measured experimentally and is
the only information that allows us to study the decay path. From the
theoretical point of view, this information is contained in the
large-distance part of the wave function, which must therefore be
computed accurately. The Zeldovic regularized Fourier transform of the
wave function gives the energy distributions\cite{fed04}. The
resonance wave functions change sometimes substantially from small to
large distances. This dynamic evolution allows a better understanding
of the decays details\cite{alv08}.

The decay mechanisms depend on the resonance properties and can be
either sequential or direct or a mixture. In our case, all of the
resonances can decay sequentially via $^8$Be($0^+$), i.e. it is
allowed by angular momentum conservation. One of the adiabatic
components is related to the $^8$Be+$n$ structure and approaches the
complex energy of the $^8$Be($0^+$) resonance. This component will give
the sequential contribution\cite{alv10}.

\subsection{$^9$Be}

\begin{figure}[th]
\vspace*{-3pt}
\centerline{\psfig{file=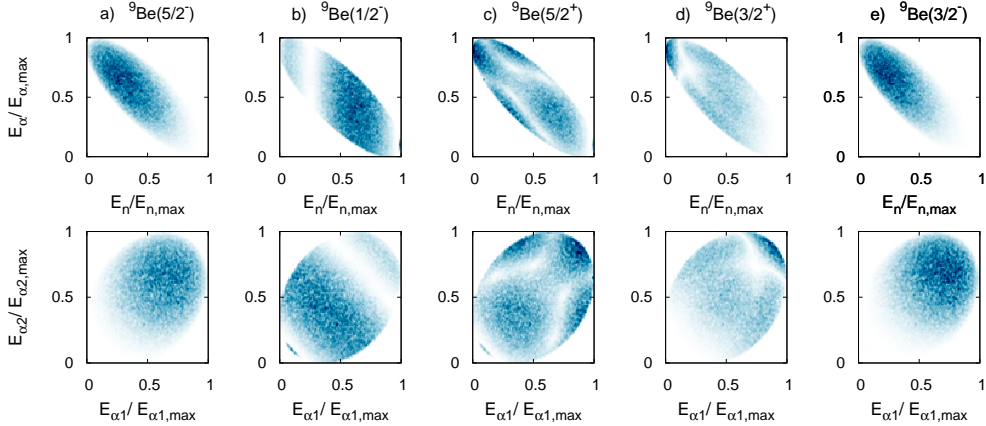,width=6.2cm,angle=270}}
\vspace*{-3pt}
\caption{Dalitz plots for a) the $5/2^-$-resonance of $^9$Be at
  2.43~MeV of excitation energy, b) the $1/2^-$-resonance of $^9$Be at
  2.82~MeV, c) the $5/2^+$-resonance of $^9$Be at 3.03~MeV, d) the
  $3/2^+$-resonance of $^9$Be at 4.69~MeV, e) the $3/2^-$-resonance of
  $^9$Be at 5.59~MeV. The upper panels show on the X-axis the neutron
  energy divided by the maximum possible, and on the Y-axis the
  $\alpha$ energy divided by the maximum possible. The lower panels
  show on X and Y the energy of the two $\alpha$-particles divided by
  the maximum. The sequential decay via $^8$Be($0^+$) has been removed.}
\label{fig9be}
\end{figure}

Fig. \ref{fig9be} corresponds to the two-dimensional energy
correlations (or Dalitz plots) of the three fragments of $^9$Be after
the decay. The two possibilities, $\alpha-n$ and $\alpha-\alpha$, are
shown and the $J^\pi$ of each state is labeled in the figure. The
energies are given in units of their maximum values for each case,
i.e. $5/9E_{res}$ for the $\alpha$-particles and $8/9E_{res}$ for the
neutrons. In all the cases we have removed the sequential decay via
$^8$Be($0^+$). It is important to remark that the results are directly
comparable to measured distributions.

The first thing that we observe in fig.~\ref{fig9be} is that all the
plots are symmetric between $\alpha$'s energies. This symmetry is
necessary since the $\alpha$'s are identical particles.  The graphs
corresponding to $\frac{5}{2}^-$ and $\frac{3}{2}^-$, are very similar
to each other. First, we do not observe zeroes in the Dalitz
plots.\cite{fyn09} This means that angular momentum conservation does
not forbid any energy combination. Second, we see that the density
increases towards higher $\alpha$ energies. This is due to the Coulomb
repulsion, which enlarges the charged particles energies while reduces
that of the neutron. The other two cases show more structure and have
zero probability points or curves.

\subsection{$^9$Be vs $^9$B}

\begin{figure}[th]
\vspace*{-3pt}
\centerline{\psfig{file=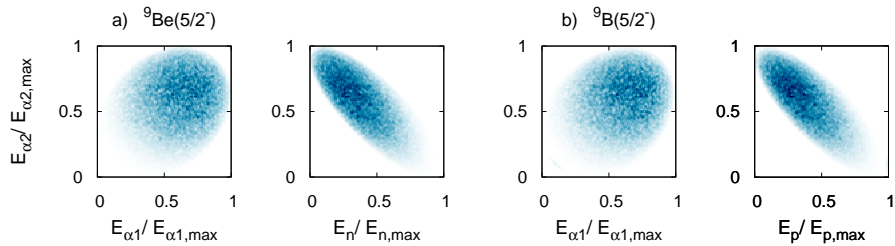,width=3.6cm,angle=270}}
\vspace*{-3pt}
\caption{Dalitz plots for a) the $5/2^-$-resonance of $^9$Be at
  2.43~MeV of excitation energy, b) the $5/2^-$-resonance of $^9$B at
  2.34~MeV. The sequential decay via $^8$Be($0^+$) has been removed.}
\label{fig9b}
\end{figure}

In our description the only difference between $^9$Be and $^9$B is the
existing Coulomb interaction between the $\alpha$-particle and the
proton in $^9$B. The structure of the resonances in both nuclei are
expected to be very similar. Fig. \ref{fig9b} shows the comparison
between the Dalitz plots corresponding to the $5/2^-$-resonance of
$^9$Be and $^9$B. The patterns are, in fact, almost indistinguishable.

\section{Summary And Conclusions}

We have described $^9$Be and $^9$B resonances as three clusters by
means of the complex-scaled hyperspherical adiabatic expansion method,
including short-range and Coulomb interactions. The two-dimensional
energy correlations of the decaying fragments are shown for the
low-lying resonances of $^9$Be. We compare one of them ($5/2^-$) to
the corresponding one in $^9$B and find, as we expected, an almost
identical distribution. Our distributions are open to experimental
tests.

\section*{Acknowledgements}

This work was partly supported by funds provided by DGI of MEC (Spain)
under the contracts FIS2008-01301 and FPA2007-62216.  R.A.R. acknowledges
support from Ministerio de Ciencia e Innovaci\'on (Spain) under the
``Juan de la Cierva'' program.

\end{document}